\documentclass[iop]{emulateapj}

\shorttitle{UV obscuration in $z \sim 2$ dusty galaxies}
\shortauthors{Penner et al.}

\begin{document}

\title{Evidence for a wide range of UV obscuration in $z \sim 2$ dusty galaxies
from the GOODS-\emph{Herschel} survey$^{\star}$}

\author{Kyle Penner\altaffilmark{1}, Mark Dickinson\altaffilmark{2},
Alexandra Pope\altaffilmark{3}, Arjun Dey\altaffilmark{2},
Benjamin Magnelli\altaffilmark{4},
Maurilio Pannella\altaffilmark{5},
Bruno Altieri\altaffilmark{6}, Herve Aussel\altaffilmark{5},
Veronique Buat\altaffilmark{7}, Shane Bussmann\altaffilmark{8},
Vassilis Charmandaris\altaffilmark{9,10,11}, Daniela Coia\altaffilmark{6},
Emanuele Daddi\altaffilmark{5}, Helmut Dannerbauer\altaffilmark{12},
David Elbaz\altaffilmark{5}, Ho Seong Hwang\altaffilmark{8},
Jeyhan Kartaltepe\altaffilmark{2}, Lihwai Lin\altaffilmark{13},
Georgios Magdis\altaffilmark{14},
Glenn Morrison\altaffilmark{15,16},
Paola Popesso\altaffilmark{4}, Douglas Scott\altaffilmark{17},
and Ivan Valtchanov\altaffilmark{6}}

\email{kpenner@as.arizona.edu}
\submitted{}
\altaffiltext{$\star$}{\emph{Herschel} is an ESA space observatory with science
instruments provided by European-led Principal Investigator consortia and with
important participation from NASA.}
\altaffiltext{1}{Department of Astronomy, University of Arizona,
Tucson, AZ 85721}
\altaffiltext{2}{National Optical Astronomy Observatory,
Tucson, AZ 85719}
\altaffiltext{3}{Department of Astronomy, University of Massachusetts,
Amherst, MA 01003}
\altaffiltext{4}{Max Planck Institut f\"ur Extraterrestrische Physik,
Postfach 1312, 85741 Garching, Germany}
\altaffiltext{5}{Laboratoire AIM Paris-Saclay,
CEA/DSM/Irfu-CNRS -- Universit\'e Paris Diderot,
CEA-Saclay, pt courrier 131, F-91191 Gif-sur-Yvette, France}
\altaffiltext{6}{Herschel Science Center, European Space Astronomy Center,
Villanueva de la Ca\~nada, 28691 Madrid, Spain}
\altaffiltext{7}{Laboratoire d'Astrophysique de Marseille, OAMP,
Universit\'e Aix-marseille, CNRS, 38 rue Fr\'ed\'eric Joliot-Curie,
13388 Marseille cedex 13, France}
\altaffiltext{8}{Harvard-Smithsonian Center for Astrophysics,
Cambridge, MA 02138}
\altaffiltext{9}{Department of Physics and Institute of Theoretical
\& Computational Physics, University of Crete,
GR-71003 Heraklion, Greece}
\altaffiltext{10}{IESL/Foundation for Research and Technology -- Hellas,
GR-71110 Heraklion, Greece}
\altaffiltext{11}{Chercheur Associ\'e, Observatoire de Paris, F-75014 Paris,
France}
\altaffiltext{12}{Universit\"at Wien, Institut f\"ur Astronomie,
T\"urkenschanzstra\ss e 17, 1180 Vienna, Austria}
\altaffiltext{13}{Institute of Astronomy \& Astrophysics,
Academia Sinica, Taipei 106, Taiwan}
\altaffiltext{14}{Department of Physics, University of Oxford,
Keble Road, Oxford OX1 3RH, UK}
\altaffiltext{15}{Institute for Astronomy, University of Hawaii,
Honolulu, HI 96822}
\altaffiltext{16}{Canada-France-Hawaii Telescope, Kamuela, HI 96743}
\altaffiltext{17}{Department of Physics \& Astronomy,
University of British Columbia, 6224 Agricultural Road,
Vancouver, BC V6T~1Z1}

\begin{abstract}
Dusty galaxies at $z \sim 2$ span a wide range of relative brightness between
rest-frame mid-infrared (8$\,\micron$) and ultraviolet wavelengths.
We attempt to determine
the physical mechanism responsible for this diversity.  Dust-obscured galaxies
(DOGs), which have rest-frame mid-IR to UV flux density ratios $\ga 1000$,
might be abnormally bright in the mid-IR, perhaps due to prominent AGN and/or
PAH emission, or abnormally faint in the UV.  We use far-infrared
data from the GOODS-\emph{Herschel} survey to show that most DOGs with
$10^{12} ~\mathrm{L}_{\odot}~ \la L_{\mathrm{IR}} \la 10^{13}
~\mathrm{L}_{\odot}$ are not
abnormally bright in the mid-IR when compared to other dusty
galaxies with similar IR (8--1000$\,\micron$) luminosities.
We observe a relation between the median IR to UV luminosity ratios
and the median UV continuum power-law indices for these galaxies, and we find
that only 24\% have specific star formation rates which indicate the
dominance of compact star-forming regions.
This circumstantial evidence supports the idea that the
UV- and IR-emitting regions in these galaxies are spatially coincident,
which implies a connection between the abnormal UV faintness of DOGs and
dust obscuration.
We conclude that the range in rest-frame mid-IR to UV flux density ratios
spanned by dusty galaxies at $z \sim 2$ is due to differing amounts of UV
obscuration.  Of galaxies with these IR luminosities, DOGs are the most
obscured.  We attribute differences in
UV obscuration to either: 1) differences in the
degree
of alignment between the spatial distributions of dust and massive stars, or 2)
differences in the total dust content.
\end{abstract}


\section{Introduction}

At $z \sim 2$, a large fraction of all high mass stars form in dusty galaxies
\citep{chapman05, magnelli11}.  Most of the intrinsic UV emission from newly
formed stars in these galaxies is obscured, or absorbed by dust grains that
subsequently
heat up and radiate at IR wavelengths.  The IR luminosity resulting from this
obscuration is usually much greater than the emergent UV luminosity.  For
galaxies in which the intrinsic UV emission from newly formed stars is less
obscured, the
IR luminosity is still greater than the emergent UV luminosity, but to a lesser
degree \citep{reddy11}.  The relation between the IR and emergent UV
emission from a $z \sim 2$ galaxy depends on the interplay between star
formation and dust obscuration.

One of the many ways to select dusty galaxies at $z \sim 2$, without
redshift determinations from spectroscopy, is to use the ratio
of observed 24 to 0.65$\,\micron$ ($R$-band) flux densities
\citep{dey08, fiore08}.  Sources
satisfying $S_{24}/S_{0.65} \ga 1000$ have been termed
``dust-obscured galaxies'', or DOGs; their redshift distribution is
approximately a Gaussian that peaks at $z = 2$ with
$\sigma_{z} = 0.5$ \citep{dey08}.  In the redshift range $1.5 < z < 2.5$,
0.65$\,\micron$ observations are sensitive to rest-frame UV emission
from newly formed massive stars, and
24$\,\micron$ observations are sensitive to mid-IR emission from hot dust and
polycyclic aromatic hydrocarbons (PAHs).  The DOG
criterion is thus
unique in that it selects galaxies in a specific redshift range, that also
exhibit extreme ratios between their rest-frame mid-IR and UV flux densities.
We have yet to understand the physical mechanism driving
the span of ratios exhibited by dusty galaxies at $z \sim 2$.

The IR luminosities of DOGs with
$L_{\mathrm{IR}} \ga 10^{13} ~\mathrm{L}_{\odot}$ are
dominated by emission from active galactic nuclei
(AGN; \citealt{dey08,bussmann09b}).  The dominant sources of the IR
luminosities of less luminous DOGs is a topic of debate.  \citet{fiore08} and
\citet{treister09} conclude that the IR luminosities of many DOGs with
$10^{12} ~\mathrm{L}_{\odot}~
\la L_{\mathrm{IR}} \la 10^{13} ~\mathrm{L}_{\odot}$ originate from AGN, while
\citet{pope08} conclude that many such DOGs are powered by newly formed stars.

In this paper, we pose the question ``What makes a DOG a DOG?''  The primary
goal of our study is determining why DOGs have an extreme ratio between
their rest-frame mid-IR and UV flux densities when compared to other dusty
galaxies.  Unfortunately, the simple and singular selection criterion cannot
distinguish between a DOG that is:
\begin{itemize}
\item abnormally bright at rest-frame $8\,\micron$ for its far-IR
flux density, indicating its mid-IR luminosity may be dominated
by AGN emission, or abnormally strong emission from polycyclic aromatic
hydrocarbons (PAHs);
\item or, abnormally faint in the rest-frame UV for its optical flux
density, indicating that dust more completely obscures the newly formed stars
in the galaxy.
\end{itemize}
We use \emph{Herschel} \citep{pilbratt10} data in the Great Observatories
Origins Deep Survey-North (GOODS-N) region \citep{elbaz11} to show that, on
average, DOGs with $10^{12} ~\mathrm{L}_{\odot}~ \la L_{\mathrm{IR}} \la
10^{13} ~\mathrm{L}_{\odot}$ are
not abnormally bright at $8\,\micron$, but
are more UV faint than other galaxies with similar IR
luminosities.
The ratio between rest-frame IR and UV flux densities is set by the amount of
obscuration,
which can vary with either: 1) the degree of
alignment
between the spatial distributions of dust and massive stars, or 2) the total
dust content.

This paper is organized as follows.  We present the data and sample selection
in \S\ref{sec:data}; in \S\ref{sec:results}, we show the results.  We discuss
the implications of these results in \S\ref{sec:discuss}, and conclude in
\S\ref{sec:conclude}.  We assume a cosmology with
$H_{0} = 70$ km s$^{-1}$ Mpc $^{-1}$, $\Omega_{\mathrm{m}} = 0.3$, and
$\Omega_{\Lambda} = 0.7$.

\section{Data}\label{sec:data}

\subsection{Measured quantities}

Our study uses observations of the GOODS-N region, which is roughly
10 arcmin $\times$ 16.5 arcmin in extent.
We cull the sample of DOGs from a catalog of 24$\,\micron$ sources
produced for the
\emph{Spitzer}/MIPS survey of the GOODS-N region (M. Dickinson, PI;
\citealt{magnelli11}).  A 24$\,\micron$ source is
defined as a $\ge 3\sigma$ flux density measurement from PSF fitting to
\emph{Spitzer}/IRAC 3.6$\,\micron$ source priors.  The catalog is
99\% complete at $S_{24} > 50\,\mu$Jy, and contains 1603 sources.

The 2.2$\,\micron$ ($K_s$-band) image we use to identify counterparts for the
24$\,\micron$ sources comes from observations
using the Canada-France-Hawaii Telescope (CFHT).  The
data are presented in \citet{wang10}; we use our own reductions \citep{lin12}.
The 0.65$\,\micron$ ($R$-band) Subaru image we use to define the DOG sample
comes from \citet{capak04}.
The 5$\sigma$ depth of the 2.2$\,\micron$ image is $\sim$0.60$\,\mu$Jy
(24.5 AB mag);
the 3$\sigma$ depth of the 0.65$\,\micron$ image is $\sim$0.05$\,\mu$Jy
(27.2 AB mag).

To extract flux densities, we follow a modified version of the procedure used
by \citet{pope08}.  Using SExtractor \citep{bertin96}, we place 3$\arcsec$
diameter apertures at the positions of sources detected ($\ge 5\sigma$) in the
2.2$\,\micron$ image.  If the 2.2$\,\micron$ flux density is detected
with S/N $\ge 5\sigma$ but the 0.65$\,\micron$ flux density is not detected
with S/N $< 3\sigma$, we use a 3$\sigma$ limit for the latter flux density.

To determine rest-frame UV continuum power-law indices, we extract flux densities
at 0.45, 0.55, 0.80, and 0.90$\,\micron$ (the $B$-, $V$-, $I$-, and $z$-bands)
from
Subaru images \citep{capak04}, using the same procedure.  We use the 3.6, 4.5,
5.8, and 8$\,\micron$ flux densities already associated with the 24$\,\micron$
sources to determine whether or not their spectral energy distributions (SEDs)
at these wavelengths behave as power laws; these flux densities come from a
catalog produced for the \emph{Spitzer}/IRAC survey of the GOODS-N region,
and will be included in catalogs accompanying the GOODS-\emph{Herschel}
public data release.

For the optical/near-IR photometry, we calculate aperture corrections, defined
as the
ratios of total flux density to flux density in a 3$\arcsec$ diameter aperture
for point sources (non-saturated stars).  We take the SExtractor parameter
\verb+FLUX_AUTO+ as the total flux density.  The corrections are factors of
1.086, 1.225, 1.247, 1.057, 1.086, and 1.057 at 0.45, 0.55, 0.65, 0.80, 0.90,
and 2.2$\,\micron$, respectively.  To maintain the signal-to-noise
ratios given by SExtractor, both the flux densities and their uncertainties
are multiplied by these factors.

We associate each 24$\,\micron$ source with a 2.2$\,\micron$ source and its
extracted optical flux densities if the 2.2$\,\micron$ source is a unique match
within 0.76$\arcsec$ of the position of the 3.6$\,\micron$ prior.  The
match
radius is chosen by maximizing the number of unique matches while minimizing
the number of multiple matches.  Of the 1603
24$\,\micron$ sources, 87 either do not have a $\ge 5\sigma$ 2.2$\,\micron$
counterpart within the match radius (85 of 87) or have multiple counterparts
(2 of 87).

The far-IR flux densities come from a catalog produced for the
GOODS-\emph{Herschel} survey \citep{elbaz11}.
We only use 100 and 160$\,\micron$ flux densities measured with PACS
\citep{poglitsch10}, and 250$\,\micron$ flux densities measured with SPIRE
\citep{griffin10}, to avoid the
complications of measuring flux densities for 24$\,\micron$ sources in the
350 and 500$\,\micron$ SPIRE images that are
affected by severe source confusion.
We consider a $\ge 3\sigma$ measurement at 100 or 160$\,\micron$ to be a
detection;
at 250$\micron$, we require a $\ge 5\sigma$ measurement.

We impose additional constraints on the 250$\,\micron$ flux densities (and
$5\sigma$ limits), similar to those defining the ``clean index''
\citep{hwang10, elbaz10, elbaz11}.  The 250$ \,\micron$ flux densities of clean
sources (and $5\sigma$ limits of clean non-detections) should
not be affected by severe source confusion.  We do not impose additional
constraints on the 100 and 160$\,\micron$ flux densities because the
100$\,\micron$ images are not deep enough to be affected by source
confusion, and the 160$\,\micron$ images are only deep enough to be moderately
affected.  The 3$\sigma$ depths of the 100 and 160$\,\micron$ images
are $\sim$1100$\,\mu$Jy and $\sim$2700$\,\mu$Jy, respectively; the 5$\sigma$
depth of the 250$\,\micron$ image is $\sim$5700$\,\mu$Jy.

We attempt to match each 24$\,\micron$ source to a source with a spectroscopic
redshift from the catalogs of \citet[][which includes redshifts compiled from
the literature]{barger08} and Stern et al. (in
preparation).  We find spectroscopic redshifts for 910 (57\%) of the
24$\,\micron$
sources.  If no coincident sources with spectroscopic redshifts are found,
we resort to the photometric redshift source catalog of Pannella et al. (in
preparation) to find a source match.  We exclude sources with photometric
redshifts derived from ill-fitting templates (those with reduced
$\chi^{2} > 2$).
For an additional 510 (32\%) of the 24$\,\micron$ sources we have photometric
redshift
estimates.  There are no redshift estimates for 183 (11\%) of the 24$\,\micron$
sources, and these sources are excluded from our samples.

\subsection{Samples}

\begin{figure}
\centering
\includegraphics[scale=0.45]{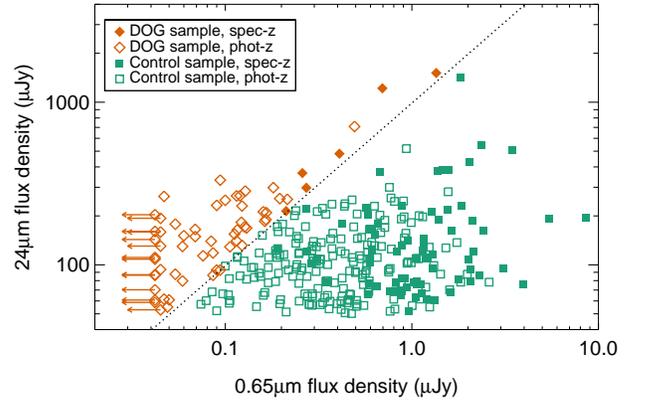}
\caption{24$\,\micron$ flux density vs. 0.65$\,\micron$ flux density for
galaxies in the two samples.  The DOG sample is defined by $S_{24}/S_{0.65} >
986$.
\label{fig:sample}}
\end{figure}

\begin{figure}
\centering
\includegraphics[scale=0.45]{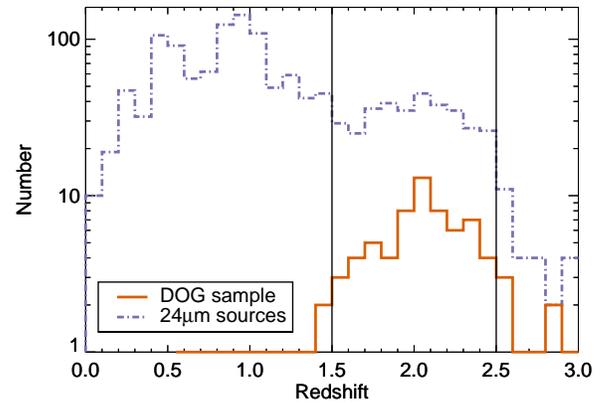}
\caption{Redshift distributions for the DOG sample and all 24$\,\micron$
sources (before we impose redshift limits).
For the analysis, we limit all samples to the redshift range
$1.5 < z < 2.5$ (the area between the vertical lines).\label{fig:z_dist}}
\end{figure}

Using the multi-wavelength information and redshifts, we define 2 samples from
the superset of all 24$\,\micron$ sources with $S_{24} > 50\,\mu$Jy:

\begin{itemize}
\item \emph{DOG sample}: All 24$\,\micron$ sources with 2.2$\,\micron$
counterparts and $S_{24}/S_{0.65} > 986$ (Fig. \ref{fig:sample}).  The
24$\,\micron$ flux density of the faintest DOG is 53$\,\mu$Jy, justifying our
$S_{24} > 50\,\mu$Jy cut for the control sample.  The limiting quantity is the
depth of the 0.65$\,\micron$ image
\citep[3$\sigma$ = 0.05$\,\mu$Jy ][]{capak04}.  The redshift distribution of
our sample
of DOGs is shown in Fig. \ref{fig:z_dist}, also motivating our $1.5 < z < 2.5$
cut for the control sample.  Six (of 61; 10\%) DOGs have spectroscopic
redshifts.  In the following analysis, we include only DOGs
with $1.5 < z < 2.5$.  Our conclusions do not change if we include DOGs without
redshift estimates in the sample.
\item \emph{Control sample}: All $S_{24} > 50\,\mu$Jy 24$\,\micron$ sources
with 2.2$\,\micron$ counterparts that are at $1.5 < z < 2.5$, and that do not
satisfy the DOG selection criterion.  Seventy four (of 268; 28\%) control
galaxies have spectroscopic redshifts.
\end{itemize}

For each sample, Table \ref{table:numbers} characterizes the subset of sources
with flux densities detected at 0.65, 100, 160, and 250$\,\micron$.
More than 70\% of these galaxies are undetected in optical spectra
(or are unobserved) because their observed-frame optical flux densities are so
faint.

Our sample contains DOGs with fainter 24$\,\micron$ emission than does the
\citet{dey08} sample.  Their sample is selected from the shallower
\emph{Spitzer}/MIPS survey of the Bo\"{o}tes region; their limit
is $S_{24} > 300\,\mu$Jy.  However, the GOODS-N region is much smaller than the
Bo\"{o}tes region, so we have few DOGs with $S_{24} > 300\,\mu$Jy.
\citet{pope08} also study a sample of DOGs
in
GOODS-N.  The main differences between the \citet{pope08} sample and ours are
that: 1) they
limit their sample to $S_{24} > 100\,\mu$Jy; and 2) they estimate a redshift
for each DOG using IRAC and MIPS photometry, whereas we match DOGs to sources
in a near-IR/optical catalog, with redshift estimates based on $UBVRIzJK$,
3.6, and 4.5$\,\micron$ photometry.

The fractions of 24$\,\micron$ sources at $1.5 < z < 2.5$ that meet the DOG
criterion increase with increasing
24$\,\micron$ flux density (Fig. \ref{fig:sample}).
Of the sources with $S_{24} < 100\,\mu$Jy, 4\% are DOGs.
Of the sources with $S_{24} > 100\,\mu$Jy, 25\% are DOGs.
\citet{riguccini11} find similar fractions; they also find that of their
sources with $S_{24} > 300\,\mu$Jy, 60\% are DOGs.

\begin{deluxetable*}{lccccccc}
\tabletypesize{\small}
\tablewidth{0pc}
\tablehead{
\colhead{} & \colhead{} & \colhead{} & \colhead{} &
\multicolumn{4}{c}{Number detected} \\
\cline{5-8} \\
\colhead{Sample} & \colhead{Number with} & \colhead{Median $z$} &
\colhead{Median $S_{24}$} & \colhead{0.65$\,\micron$} &
\colhead{100$\,\micron$} & \colhead{160$\micron$} & \colhead{250$\,\micron$} \\
 & $1.5 < z < 2.5$ & & $\mu$Jy & $\ge 3\sigma$ & $\ge 3\sigma$ &
$\ge 3\sigma$ &
$\ge 5\sigma$, clean}
\startdata
DOGs            &     61 & 2.1 & 161 & 47 (77\%) &           29 (48\%) & 24
(39\%) &          9 (15\%) \\
Control         &    268 & 2.0 & 102 & 268 (100\%) &           81 (30\%) & 52
(19\%) &          15 (6\%)
\enddata

\tablecomments{All sources have $\ge 5\sigma$ 2.2$\,\micron$ and $\ge 3\sigma$
24$\,\micron$
flux density measurements.  We also require $S_{24} > 50\,\mu$Jy.
\label{table:numbers}}
\end{deluxetable*}

\subsection{Derived quantities}

Several quantities are useful in analyzing the relation between IR and
emergent UV emission from galaxies.  In this section, we detail how we
estimate the total IR and UV
luminosities, UV continuum power-law indices, star formation rates, and stellar masses for
the galaxies in our samples.

\subsubsection{IR luminosities}

We estimate a total IR luminosity (8--1000$\,\micron$; $L_{\mathrm{IR}}$) for
each galaxy with detected emission
at 100$\,\micron$.  We redshift the \citet{chary01} template spectral energy
distributions (SEDs) to the distance of each galaxy, find the SED that most
closely matches the observed 100$\,\micron$ flux density, and multiply the
IR luminosity of that SED by the ratio between actual and predicted flux
densities to get $L_{\mathrm{IR}}$.

We prefer this approach over estimating the IR luminosity directly, by
summing several far-IR flux densities, because the
latter procedure requires detected emission at 160 and 250$\,\micron$.  The
GOODS-\emph{Herschel} image at 160$\,\micron$ is moderately affected by
blending due to source confusion, while the image at 250$\,\micron$ is so deep
that blending is problematic.  The drawback to our chosen approach is
that all statements we make regarding IR luminosities assume that the
low-redshift template SEDs accurately represent the SEDs of galaxies at
$z \sim 2$.  \citet{elbaz10} show that this assumption is valid when template
matching is done to 100$\,\micron$ flux densities.

\subsubsection{UV continuum power-law indices and luminosities}

For galaxies with UV emission from newly formed massive stars, the
UV continuum can be approximated as a power law with an index $\beta$:
\begin{equation}
S_{\lambda} = C\lambda^{\beta}.
\end{equation}
We use $\ge 3\sigma$ flux densities at 0.45, 0.55, 0.65, 0.80, and
0.90$\,\micron$ (i.e., in the $B$-, $V$-, $R$-, $I$-, and $z$-bands) to fit for
$\beta$ and the constant factor $C$ for each galaxy which has an estimate of
$L_{\mathrm{IR}}$.  If only two flux densities have S/N $\ge 3\sigma$, we
calculate $\beta$ analytically.  If only one flux density has S/N
$\ge 3\sigma$, we use an upper limit for the 0.45$\,\micron$ flux density to
calculate a lower limit to $\beta$.

We estimate a UV luminosity $\lambda L_{\lambda}$ at rest-frame 0.16$\,\micron$
for each galaxy using its redshift estimate and the power-law fit to the
rest-frame UV flux densities.

\subsubsection{Stellar masses and star formation rates}

To estimate a stellar mass for each galaxy, we fit stellar population synthesis
models to its $UBVRIzJK$, 3.6, and 4.5$\,\micron$ flux densities
\citep{drory04, pannella09b}.  Full details are in \S4.2 of \citet{mullaney12}.
We assume
the stellar initial mass function in \citet{salpeter55} from 0.1 to
100~$M_{\odot}$, as well as the dust attenuation law in \citet{calzetti00}.

We use the equation in \citet{kennicutt98} to calculate a star formation rate
(SFR), based on the IR luminosity, for each galaxy with detected 100$\,\micron$
emission.  We make no correction for the emergent UV luminosity, since it is
negligible for all galaxies in the DOG and control samples
(see \S\ref{sec:discuss}).  In using
the \citet{kennicutt98} equation, we assume that the observed 100$\,\micron$
emission is due to star formation and not AGN activity, and that the star
formation episode lasts for $< 10^{8}$ years.  \citet{mullaney12} find that
the former assumption is valid for most AGN with detected X-ray emission
at $z < 3$.

\section{Results}\label{sec:results}

\begin{figure}
\centering
\includegraphics[scale=0.45]{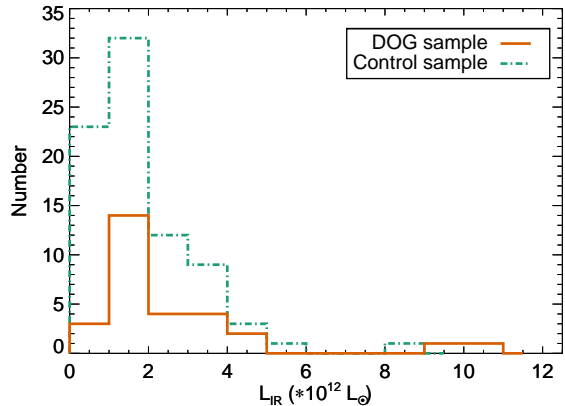}
\caption{Distributions of IR luminosities (derived from best-fit templates)
for galaxies with
detected emission at 100$\,\micron$.  The two distributions are statistically
indistinguishable.
\label{fig:lir}}
\end{figure}

Seventy six percent of the $z \sim 2$ galaxies with detected
100$\,\micron$
emission
have $10^{12} ~\mathrm{L}_{\odot}~ \la L_{\mathrm{IR}} \la 10^{13}
~\mathrm{L}_{\odot}$ (Fig. \ref{fig:lir}).  The distributions of IR
luminosities for the DOGs and the
control galaxies are statistically indistinguishable ($p = 0.20$ that the
two samples are drawn from the same parent population, using a K-S test).

\begin{figure}
\centering
\includegraphics[scale=0.45]{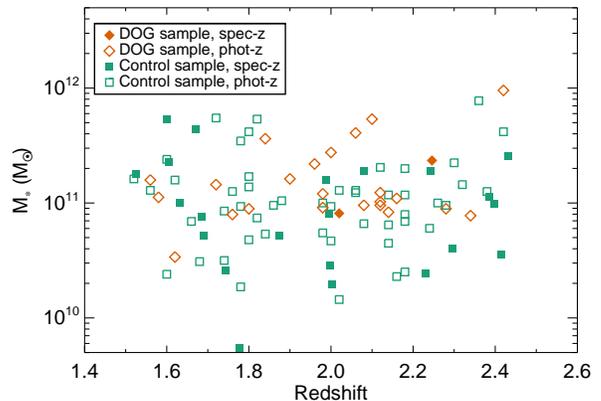}
\caption{Stellar mass vs. redshift for galaxies with
detected emission at 100$\,\micron$.  The distribution of stellar masses for
DOGs is not strongly different from that for control galaxies, though
some control galaxies have lower stellar masses than do the DOGs.
Twenty four
percent of the control sample has $M_{\ast} <
5\times10^{10} \mathrm{M}_{\odot}$, as opposed to 4\% of the DOG
sample.\label{fig:mass_z}}
\end{figure}

Fig.~\ref{fig:mass_z} shows the stellar masses for DOGs and control galaxies
with detected emission at 100$\,\micron$.  The distributions are not strongly
different ($p = 0.02$, using a K-S test), though some control galaxies
have lower stellar masses than do the DOGs.  17 (of 72; 24\%) control galaxies
have stellar masses $M_{\ast} < 5\times10^{10} \mathrm{M}_{\odot}$,
while only 1 (of 25; 4\%) DOG has a stellar mass below this threshold.

The infrared SEDs of most low redshift dusty galaxies peak between rest-frame
$\sim 60$ and 100$\,\micron$.  The rest-frame $8\,\micron$ luminosity is only a
fraction of the total IR luminosity in these galaxies.  To address whether or
not DOGs are abnormally bright at rest-frame $8\,\micron$, we
require a comparison of the rest-frame far-IR and $8\,\micron$ flux
densities between DOGs and the control galaxies.

\begin{figure}
\centering
\includegraphics[scale=0.45]{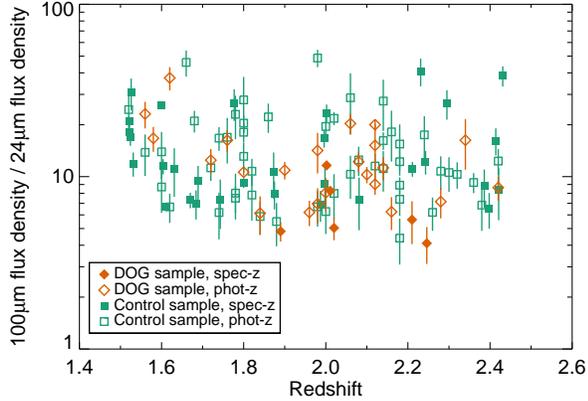}
\caption{$S_{100}/S_{24}$ vs. redshift for galaxies in the two samples.
Galaxies with limits at 100$\,\micron$ are not shown in this figure, but we do
include them in our statistical tests.  DOGs are statistically
indistinguishable from the control galaxies in $S_{100}/S_{24}$.
\label{fig:100over24_z}}
\end{figure}

Fig. \ref{fig:100over24_z} shows that DOGs are statistically indistinguishable
from the control galaxies when looking at the observed flux density ratio
$S_{100}/S_{24}$ ($p = 0.35$, using a K-S test) .  In this figure, we do not
show the galaxies with limits for
$S_{100}$.  We perform the Gehan and logrank tests, which are
conceptually similar to the two sample K-S test, but also allow for the
inclusion of galaxies with limits at 100$\,\micron$ in the two samples.  We
find no statistically significant difference between the DOGs and the control
galaxies ($p > 0.70$ that both samples are drawn from the same parent
population, from both tests).  Our conclusions are the same using
$S_{160}/S_{24}$ (not shown; $p > 0.10$ from both tests).  If DOGs were
abnormally luminous at rest-frame
$8\,\micron$ for their far-IR luminosities, then we would expect them to have
low values of observed $S_{100}/S_{24}$ and/or $S_{160}/S_{24}$ compared to
those of the control galaxies; they do not.  However, 100 and 160$\,\micron$
observations are sensitive to rest-frame 33 and 53$\,\micron$ emission from
galaxies at $z = 2$.  These rest-frame wavelengths are still short of the
presumed wavelength of the peak of the infrared SED; the
rest-frame luminosities are still only a fraction of the total IR
luminosity in these galaxies.

\begin{figure}
\centering
\includegraphics[scale=0.45]{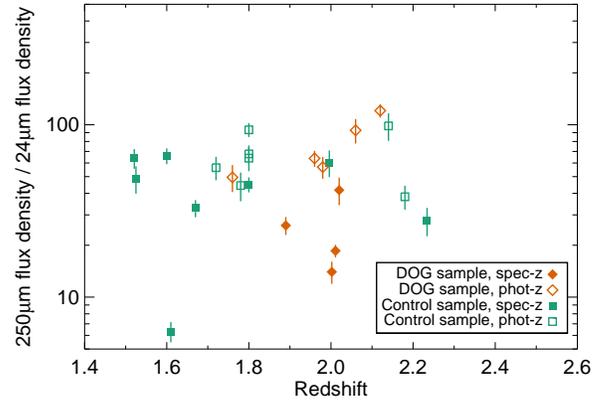}
\caption{$S_{250}/S_{24}$ vs. redshift for galaxies in the two samples.
Galaxies with limits at 250$\,\micron$ are not shown in this figure, but we do
include them in our statistical tests.  As in Fig.
\ref{fig:100over24_z}, DOGs span the same range of flux density ratios as
do the control
galaxies; this leads us to conclude that the DOG criterion does not select
dusty galaxies that are abnormally bright at rest-frame $8\,\micron$.
\label{fig:250over24_z}}
\end{figure}

250$\,\micron$ observations are sensitive to rest-frame $83\,\micron$ emission
from galaxies at $z = 2$.  This rest-frame wavelength is generally close to
the wavelength of the peak of the IR SED.
Fig. \ref{fig:250over24_z} shows that DOGs are statistically indistinguishable
from the control galaxies when looking at the observed flux density ratio
$S_{250}/S_{24}$.  We reach the same conclusion when including
galaxies with limits at 250$\,\micron$ in statistical tests ($p > 0.76$ that
both samples are drawn from the same parent population, from both a Gehan and
logrank test).

\begin{figure}
\centering
\includegraphics[scale=0.45]{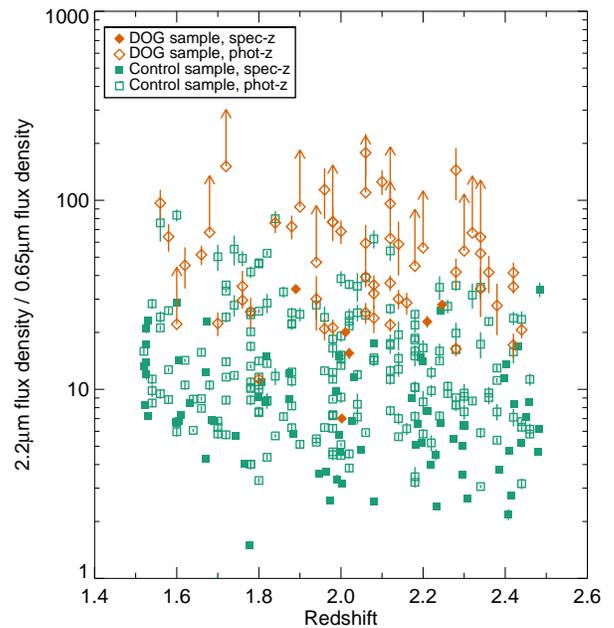}
\caption{$S_{2.2}/S_{0.65}$ vs. redshift for galaxies in the two samples.
A DOG is distinguishable from a random galaxy with detected 24$\,\micron$
emission because it is abnormally faint in the rest-frame UV.
\label{fig:RoverK_z}}
\end{figure}

The DOG criterion does not select galaxies that are abnormally bright at
rest-frame $8\,\micron$ for their far-IR flux densities.  What makes a DOG a
DOG must be that the galaxy's rest-frame UV emission is suppressed, compared to
the UV emission from a control galaxy.  Therefore we expect
a clear separation between DOGs and control galaxies when looking at the
ratios of rest-frame optical to UV flux densities.  Indeed, Fig.
\ref{fig:RoverK_z} shows that 92\% of DOGs have an observed
$S_{2.2}/S_{0.65} > 20$, while 78\% of control galaxies have
$S_{2.2}/S_{0.65} \la 20$.  The Gehan and logrank tests report a statistical
difference between the DOGs and control galaxies ($p < 0.0001$ that both
samples are drawn from the same parent population, from both tests).  Galaxies
with $S_{2.2}/S_{0.65} \ga 20$ are also known as ``extremely red objects'', or
EROs (\citealt{elston88}; \citealt{hu94}; \citealt{graham96};
\citealt{mccarthy04}).  We note that even though a galaxy with detected
24$\,\micron$ emission may meet the ERO criterion, it may not meet the DOG
criterion (and vice versa, rarely).

The statistical difference between the DOGs and control galaxies 
for $S_{2.2}/S_{0.65}$, and the lack thereof for $S_{100}/S_{24}$,
is robust against photometric redshift errors.  We offset
photometric redshifts for the superset of all galaxies with detected
24$\,\micron$ emission, assuming the offsets obey a Gaussian
distribution with $\sigma_{\Delta z/(1+z)}$ = 0.1.  We then select new DOG
and control samples.  The differences in $S_{2.2}/S_{0.65}$ are always
statistically significant, and the differences in $S_{100}/S_{24}$ are very
rarely statistically significant (from K-S tests).

\section{Discussion}\label{sec:discuss}

We have shown that the 100, 160, and 250 to 24$\,\micron$ flux density
ratios for DOGs with moderate IR luminosities
($10^{12} ~\mathrm{L}_{\mathrm{\odot}}~
< L_{\mathrm{IR}} < 10^{13} ~\mathrm{L}_{\mathrm{\odot}}$) are statistically
indistinguishable from those flux density ratios for galaxies with detected
24$\,\mu$m emission that lie at similar redshifts and have similar IR
luminosities, but that do not meet the DOG selection criterion.  Most DOGs have
higher 2.2 to 0.65$\,\micron$ flux density ratios than do the control galaxies.
Thus it seems clear that DOGs occupy the tail of a distribution of UV
obscuration in IR-luminous galaxies at $z \sim 2$.

We select a sample of DOGs using deep images of the GOODS-N
region.  These DOGs have lower IR luminosities than do DOGs selected using
shallower images of wider regions.  For example, the \citet{dey08} sample in
the wide Bo\"{o}tes region contains many DOGs with
$L_{\mathrm{IR}} > 10^{13} ~\mathrm{L}_{\odot}$.  Our sample contains very few
galaxies with $L_{\mathrm{IR}} > 10^{13} ~\mathrm{L}_{\odot}$, and our
conclusions regarding obscuration may not apply to DOGs with such high
luminosities.  At low redshift, AGN emission dominates the IR
luminosity for most galaxies with
$L_{\mathrm{IR}} > 10^{13} ~\mathrm{L}_{\odot}$
\citep{tran01}.  \citet{dey08} find that many DOGs at $z \sim 2$ with
$L_{\mathrm{IR}} > 10^{13} ~\mathrm{L}_{\odot}$ have featureless SEDs from
observed-frame 3.6 to 8$\,\micron$, indicating the presence of AGN-heated
dust.
Several studies \citep{tyler09,bussmann09b,melbourne12} find that, for these
DOGs, the SEDs at rest-frame mid- and far-IR wavelengths are
similar to those of low redshift galaxies with IR luminosities dominated by
AGN emission, such as Markarian 231.  Only 2 (of 59; 3\%) DOGs,
and 8 (of 253; 3\%) control galaxies, have
SEDs that increase (in $\nu S_{\nu}$) from observed-frame 3.6 to 8$\,\micron$,
and indeed both are among the DOGs with the highest $L_{\mathrm{IR}}$
(at $4.8\times10^{12}$ and $9.5\times10^{12} ~\mathrm{L}_{\odot}$).
Two galaxies are not enough to allow us to rule out the possibility that DOGs
with $L_{\mathrm{IR}} > 10^{13} ~\mathrm{L}_{\odot}~$ differ in their 100 or
250 to 24$\,\micron$ flux density ratios from control galaxies with similar
luminosities.

Most galaxies in our sample have $10^{12} ~\mathrm{L}_{\mathrm{\odot}}~
< L_{\mathrm{IR}} < 10^{13} ~\mathrm{L}_{\mathrm{\odot}}$.  Whether their
IR luminosities are dominated by emission from AGN or newly formed stars
is not clear.  \citet{fiore08} and \citet{treister09} find that the
average DOG in this $L_{\mathrm{IR}}$ range has an X-ray spectrum with a
power-law index they interpret as indicative of heavily obscured AGN emission.
An obscured AGN would presumably contribute to the IR luminosity as well.
\citet{pope08} find the same average X-ray index for a similar sample of DOGs,
but also find PAH emission in the mid-infrared spectra of the DOGs; they
conclude that heavily obscured AGN emission cannot coexist with PAH emission.
\citet{pope08} attribute the X-ray emission of the DOGs to X-ray binaries
rather than AGN.  In our sample, 5
(of 61; 8\%) DOGs and 31 (of 268; 12\%) control galaxies have detected X-ray
emission \citep{alexander03}.  For galaxies with
$L_{\mathrm{IR}} < 10^{13} ~\mathrm{L}_{\odot}$, we do not find any
evidence that AGN emission at rest-frame mid-IR wavelengths is more or less
common in DOGs than in the control galaxies: the fractions of DOGs and
control galaxies with increasing SEDs between observed-frame 3.6$\,\micron$
and 8$\,\micron$ are similar, as are the fractions with detected X-ray
emission.  No matter what powers
their IR luminosities, DOGs in this $L_{\mathrm{IR}}$ range are abnormally UV
faint, and are probably the most obscured galaxies.

We have shown that the brightness of the emergent UV emission in
$z \sim 2$ dusty galaxies can vary considerably from galaxy to galaxy, even
though the emergent UV emission is only a fraction of the IR emission.
In a dusty galaxy at either low or high redshift, most of the intrinsic UV
emission from
newly formed stars and/or AGN is absorbed by dust grains, which emit at IR
wavelengths.  We now ask: ``Are the UV-emitting regions that we see from a DOG
or a control galaxy spatially coincident with the IR-emitting regions?''  If
yes, the variation in emergent UV emission is due to UV obscuration.  If no,
the emergent UV emission may come from stars in a ``normal'' galactic disk,
whereas the newly formed stars ultimately responsible for the IR emission are
completely obscured, perhaps in a compact, nuclear star-forming region.  In
this scenario,
the variation in the rest-frame optical to UV flux density ratios might be due
as much to differences in the stellar populations from galaxy to galaxy as to
differences in obscuration \citep{charmandaris04}.

We cannot yet efficiently spatially resolve the IR-emitting regions for large
samples
of high redshift dusty galaxies.  Two pieces of circumstantial evidence support
the statement
that the UV- and IR-emitting regions in these dusty galaxies are
spatially coincident: 1) a minority of DOGs and control galaxies deviate
from the median specific SFR for galaxies at $z \sim 2$; and 2)
DOGs and control galaxies define a continuous relation between their
median IR to UV luminosity ratios and median UV continuum power-law indices.

\begin{figure}
\centering
\includegraphics[scale=0.45]{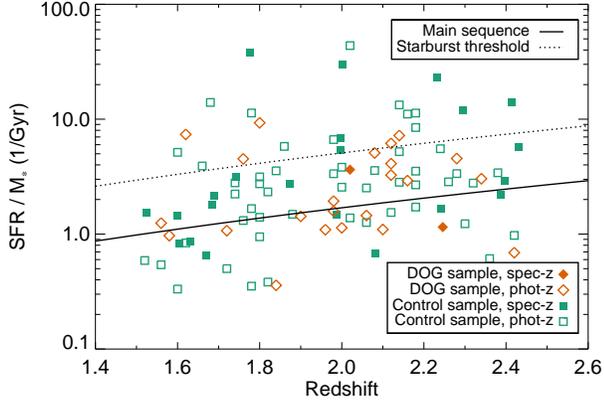}
\caption{Specific star formation rate (SFR over stellar mass)
vs. redshift
for galaxies with detected 100$\,\micron$ emission.  The solid line is the
median specific SFR for star-forming galaxies, as a function of redshift
\citep{elbaz11}.  The dotted line is three times the median value, which
we use as a threshold for identifying starbursts. 20\% of DOGs and 25\% of
control galaxies are starbursts.  DOGs deviate
from the median specific SFR no more frequently than do the control galaxies.
\label{fig:mainseq}}
\end{figure}

The specific star formation rate of a galaxy is the ratio of its SFR to its
stellar mass.  Most star-forming galaxies at the same redshift
fall on a tight relation between SFR and stellar mass; this relation is
referred to as the
``main sequence'' \citep{noeske07,elbaz07,daddi07,pannella09a}. \citet{elbaz11}
find that
SFR and stellar mass are directly proportional at all redshifts, and present
the evolution of the median specific SFR for galaxies at $0 < z < 2.5$.
\citet{elbaz11} also isolate the minority of galaxies with
specific SFRs much higher than the median values
(``starbursts'').  Here, we define starbursts as galaxies
with specific SFRs higher than 3 times the median value at their
redshift.  A minority of DOGs (5 of 25; 20\%) and control galaxies
(18 of 72; 25\%) are starbursts (Fig. \ref{fig:mainseq}).  \citet{elbaz11}
find that individual low redshift starbursts have compact UV-emitting regions;
the correspondence between deviation from the median specific SFR and
the compact size of the UV-emitting region also holds for the average
$z \sim 2$ starburst.
If this correspondence holds for individual $z \sim 2$ galaxies, few
DOGs and control galaxies would have concentrated UV-emitting regions.
These regions in the typical DOG or control galaxy would be more widely
distributed, and it would be less plausible that the UV-emitting regions
occupy parts of the galaxy not occupied by IR-emitting regions.
That radio-emitting regions in high redshift galaxies are widely
distributed indirectly supports this idea \citep{rujopakarn11}.

For starburst galaxies at low redshift, \citeauthor{meurer99}
\citep[\citeyear{meurer99}, and more recently][]{overzier11} find a relation
between the ratio of IR to UV
luminosities (denoted IRX) and the power-law index of the SED in the UV (denoted
$\beta$).  (\citealt{meurer99} refer to low redshift galaxies with
bright UV-emitting regions as ``starbursts''; the term does not necessarily
refer to galaxies that deviate from the main sequence.)  This relation is
generally interpreted to mean that the IR emission
originates as UV emission from newly formed
stars, which is partially absorbed by dust.  The dependence between the
emergent UV emission, the dust emission in the IR, and $\beta$, in
the local IRX-$\beta$ relation is consistent with the dependence of dust
absorption on $\beta$ in the
\citet{calzetti94,calzetti00} dust attenuation law.  In other words, we think
that
galaxies that lie on this relation are thought to have spatially coincident
UV- and IR-emitting regions.  \citet{reddy10,reddy11} show that most
Lyman break galaxies (LBG) at $z \sim 2$ lie on the local IRX-$\beta$ relation.
However,
both $z \sim 2$ LBGs and the low redshift starbursts in
\citet{meurer99} are much less dusty than DOGs. Some sub-classes of dusty
galaxies deviate from the local relation, perhaps because their IR emission is
unrelated to their emergent UV emission
\citep{goldader02, chapman05, papovich06, bauer11}.

\begin{figure}
\centering
\includegraphics[scale=0.45]{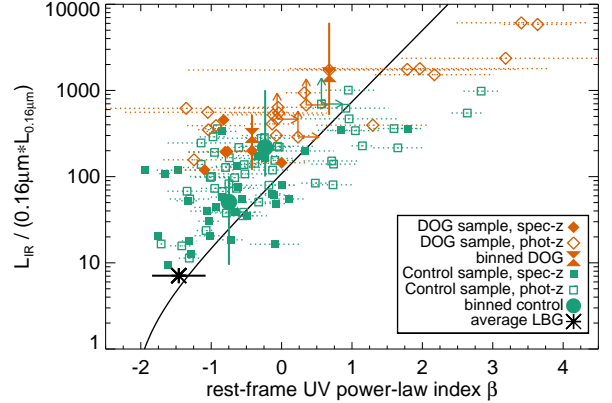}
\caption{$L_{\mathrm{IR}}$ over 0.16$\,\micron$ luminosity (IRX) vs. UV
continuum power-law index ($\beta$) for galaxies with detected 100$\,\micron$ emission.
We only show errors
on the UV power-law indices.  Filled hourglasses and circles show
the median $\beta$ values, in bins of IRX; the vertical bars show the range of
IRX over which the $\beta$ values are binned.  The
``average LBG'' black asterisk is from \citet{reddy11}, and shows a mean and
its dispersion.  The solid line is the local IRX-$\beta$ relation from
\citet{overzier11}.
\label{fig:irxbeta}}
\end{figure}

Fig.~\ref{fig:irxbeta} shows the local IRX-$\beta$ relation from
\citet{overzier11} and the relevant quantities
for the $z \sim 2$ dusty galaxies in our samples.   Individual
galaxies, in both the DOG and control samples, are found on either side of the
relation.
Because DOGs are so faint in the UV, it is difficult to precisely
determine the power-law indices of their UV SEDs; the large uncertainties on $\beta$
values preclude us from
reaching firm conclusions about individual galaxies.
The median $\beta$ values
for the DOG and control samples tend to lie to the left of the local
IRX-$\beta$ relation; DOGs do not appear to deviate more strongly
from the relation than do control galaxies.  These $z \sim 2$ dusty galaxies
define their own IRX-$\beta$ relation.  As noted earlier, specific
sub-populations of dusty galaxies deviate from the local relation, but the
trend that we find between the averages for a broader $z \sim 2$ population
has not been discussed.  Because the median $\beta$
value increases with increasing IRX, from the LBGs to the DOGs, the IR emission
is not completely independent of the emergent UV emission in these galaxies.
This reinforces our initial conclusion that $z \sim 2$
dusty galaxies populate a continuum of UV obscuration, and that DOGs are
simply the most heavily obscured galaxies.

DOGs in the luminosity range spanned by our sample are more obscured than the
control galaxies for a reason unrelated to their far-IR luminosity, because
the $L_{\mathrm{IR}}$ distributions of the two samples are statistically
indistinguishable.  In galaxies with dust
heated by UV emission from newly formed stars, the amount of obscuration
affecting that UV emission can vary due to either:
1)
differences in the degree of alignment between the spatial distributions of
dust and newly formed massive stars; or 2) differences in the total dust
content.

\citet{meurer99} derive the local
IRX-$\beta$ relation by assuming a uniform screen of dust in the line of sight
between us and the newly formed stars (see figure 8 in \citealt{calzetti94} for
an illustration).  \citet{gordon97} show that this assumption can be recovered
when the dust is instead distributed in clumps around star-forming regions.
Thus, DOGs might be more obscured than the control galaxies because DOGs have
more dust clumps surrounding star-forming regions.

Variations in UV obscuration might also be caused by
differences in the physical properties responsible for the shape of the
sub-millimeter SED.  For example, increases in the mass of cold dust can result
in greater UV obscuration.  Increases in the cold dust mass can also result in
increases in sub-millimeter emission.  Since the
sub-millimeter luminosity is usually a small fraction of the total IR
luminosity, the increase in total IR luminosity due to a higher cold dust mass
would not be easily measurable between 100 and 250$\,\micron$.
We cannot fully test this
hypothesis without sub-millimeter luminosities for galaxies in our samples.
The sensitivities of current sub-millimeter facilities limit
comparisons of the dust content between DOGs and control galaxies to the most
luminous  ($L_{\mathrm{IR}} > 10^{13} ~\mathrm{L}_{\odot}$) galaxies
\citep{bussmann09b, magdis11}.

Currently available data do not allow us to discriminate between any
hypotheses for the physical mechanisms
responsible for either the differences in the degree of alignment between
dust and stars or
differences in the dust content.  For instance, two plausible hypotheses
are that: 1) galaxy inclination might be responsible for the patchiness of
dust in the line-of-sight, and 2) many DOGs are merging galaxies, and some
aspect of the merging process creates differences in the degree of alignment.
To falsify either we require high resolution rest-frame optical
images; the Cosmic Assembly Near-infrared Deep Extragalactic Legacy Survey
(CANDELS) will obtain such images using \emph{HST}/WFC3 in the near future
\citep{grogin11}.  We note that \citet{kartaltepe12} and \citet{schawinski12}
examine rest-frame optical images of DOGs in GOODS-S and conclude that
most are undisturbed disk galaxies.  Furthermore, \citet{narayanan10}
simulate isolated galaxies that meet the DOG selection criterion, so DOGs are
not necessarily merging galaxies on the basis of their extreme rest-frame
mid-IR to UV flux density ratios.

Finally, UV obscuration is affected by the prominence of the ``dust bump'' at
rest-frame 2175~\AA~in a galaxy's attenuation curve.  At $z \sim 2$, rest-frame
2175~\AA~is redshifted into the $R$-band filter.  The presence of this
feature in the attenuation curves of high redshift galaxies has been
controversial; the bump is found, though, in the attenuation curves for several
samples of
galaxies at $z > 1$ \citep{noll09, buat11, buat12}.  However, doubling the
amplitude of
the average bump leads to a reduction in the 0.65$\,\micron$ flux density by a
factor of 1.3.  This is not large enough to explain the spread of
$S_{2.2}/S_{0.65}$ in Fig. \ref{fig:RoverK_z}.

\section{Conclusions}\label{sec:conclude}

\begin{enumerate}
\item We cull a sample of dust-obscured galaxies (DOGs), or galaxies with
$S_{24}/S_{0.65} > 986$ that are at $1.5 < z < 2.5$, in the GOODS-N region.
We use deep GOODS-\emph{Herschel} data to compare the emission from DOGs with
that from other $z \sim 2$ galaxies with detected 24$\,\micron$ emission.
\item The DOGs in our sample span
$10^{12} ~\mathrm{L}_{\odot}~ \la L_{\mathrm{IR}} \la 10^{13}
~\mathrm{L}_{\odot}$.  DOGs and control galaxies, with detected 100$\,\micron$
emission, have similar distributions of IR luminosities and stellar masses.
\item We compare the rest-frame far-IR and optical flux densities of DOGs with
those of the control galaxies.  DOGs have extreme ratios of $S_{24}/S_{0.65}$
not because they are abnormally bright at rest-frame $8\,\micron$ for their
far-IR flux densities, but because they are abnormally faint in the rest-frame
UV.
\item DOGs and the control galaxies scatter around the median
specific SFR established by $z \sim 2$ galaxies falling on the
``main sequence''; 20\% of DOGs have specific SFRs greater
than 3 times the median value, and are thus starbursts.
If UV-emitting regions in high-redshift starbursts are distributed as they
are in low-redshift starbursts, few DOGs have compact
UV-emitting regions.
\item For both the DOG and control samples, the median rest-frame UV continuum
power-law index ($\beta$) at a given IR to UV luminosity ratio (IRX) is lower than
the index predicted by the local IRX-$\beta$ relation.  DOGs do not
appear to deviate more from this relation than do the control galaxies.
Over more than a factor of 100 in IRX, the median $\beta$ value for these
galaxies systematically increases with increasing IRX.
\item These pieces of evidence suggest that, for most of these
galaxies,
the UV- and IR-emitting regions are spatially coincident.  Thus, the range in
rest-frame mid-IR to UV flux density ratios spanned by dusty galaxies at
$z \sim 2$ is due to differing amounts of UV obscuration.
DOGs are the most heavily obscured galaxies.
\item Differences in the amount of obscuration between DOGs and other dusty
galaxies at $z \sim 2$ can be due to: 1)
differences
in the degree of alignment between the spatial distributions of dust and
massive stars, or 2) differences in the total dust content.
\end{enumerate}

Our samples do not have many galaxies with
$L_{\mathrm{IR}} \ga 10^{13} ~\mathrm{L}_{\odot}$.  In DOGs with these IR
luminosities, where AGN emission may be the dominant source of dust heating,
our conclusions about obscuration may not be valid.

Further information about the nature of obscuration in these galaxies can come
from measurements that spatially resolve UV- and IR-emitting regions.  For
instance, the H$\alpha$ to H$\beta$ line ratio varies with the amount of dust
absorption, so its variation across a galaxy traces the spatial dust
distribution.  \citet{brand07} measure the galaxy-wide Balmer decrement for
several very luminous DOGs (those most likely to have IR
luminosities dominated by AGN emission);  \citet{melbourne11} present
spatially resolved H$\alpha$ maps, but H$\beta$ is outside their spectral
window.  \emph{HST} can
resolve the rest-frame UV and optical emission from these galaxies
\citep{bussmann09a,bussmann11,kartaltepe12}, and ALMA will resolve the IR
emission and allow us to measure sub-millimeter luminosities we need to
determine cold dust mass.  Studies using such data will clearly
establish the degree to which the spatial distributions of dust and massive
stars align.

\acknowledgements This work is based on observations made with \emph{Herschel},
a European Space Agency Cornerstone Mission with significant participation by
NASA. Support for this work was provided by NASA through an award issued by
JPL/Caltech.  The research activities of M.D. and A.D. are supported by NOAO,
which is operated by the Association of Universities for Research in Astronomy
under a cooperative agreement with the National Science Foundation.
We are grateful to N. Drory for sharing the SED-fitting code we use to estimate
galaxy stellar masses.

\end{document}